\documentstyle[aps,prl,floats,twocolumn,epsf]{revtex}

\begin{document}


\wideabs{
\title{Evidence for an Anisotropic State of Two-Dimensional Electrons
in High Landau Levels}

\author{M.~P. Lilly$^1$, K.~B.~Cooper$^1$, J.~P. Eisenstein$^1$, 
L.~N. Pfeiffer$^2$, and K. W. West$^2$}

\address{$^1$California Institute of Technology, Pasadena CA 91125 \\
         $^2$Bell Laboratories, Lucent Technologies, Murray Hill, NJ 07974}

\maketitle

\begin{abstract}
Magneto-transport experiments on high mobility two-dimensional electron
gases in GaAs/AlGaAs heterostructures have revealed striking anomalies
near half-filling of several spin-resolved, yet highly excited, Landau
levels. These anomalies include strong anisotropies and non-linearities
of the longitudinal resistivity $\rho_{xx}$ which commence only below
about 150mK. These phenomena are not seen in the ground or first excited
Landau level but begin abruptly in the third level. Although their
origin remains unclear, we speculate that they reflect the spontaneous
development of a generic anisotropic many-electron state. 
\end{abstract}

}

A magnetic field applied perpendicular to the plane of a
two-dimensional electron gas (2DEG) resolves the energy spectrum
into discrete Landau levels (LLs). As the field increases, the Fermi
level drops down through the Landau ladder in a series of 
steps until, at high field, it resides in the lowest (N=0) level.
In this situation the kinetic energy of the electrons is 
quenched and electron-electron interactions dominate the physics with
the fractional quantized Hall effect (FQHE) as the most spectacular
consequence\cite{review}. After more than 15 years of study, much is
known about electron correlations in this lowest LL case. The same
cannot be said when the Fermi level is in a higher Landau level. In the
second LL (N=1), the FQHE is virtually absent; only fragile and poorly
understood states at Landau filling fractions $\nu$=7/3, 5/2, and 8/3
are seen in the best samples. In the third and higher LLs (N$\ge$2)
still less is known, although there
have been interesting suggestions of charge density
waves in the clean limit\cite{CDW1,CDW2}. At very high N,
and therefore very low magnetic field, the Landau level splitting
becomes insignificant and the 2DEG assumes the character of a weakly
disordered Fermi liquid.

In this paper we report the observation of several dramatic anomalies in
the low temperature magneto-transport of clean 2DEGs when the
Fermi level lies near the middle of a spin-resolved highly excited
Landau level. These effects, which commence only below about 150mK,
abruptly begin and are strongest in the
third (N=2) LL, but persist up to about N=6. Including strong 
anisotropies and intriguing non-linearities of the
resistivity, these effects
suggest a considerably more interesting tableau at high N than
independent electrons moving in a disordered Landau band.

\begin{figure}[t]
\begin{center}
\epsfxsize=3.3in
\epsffile[38 158 534 500]{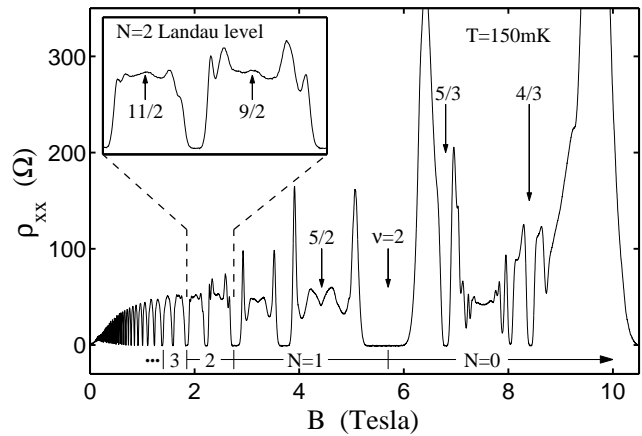}
\end{center}
\caption[figure 1]{Overview of diagonal resistivity in sample A at
T=150mK. Structure in the N=2 Landau level is expanded in the inset.}
\end{figure}

The samples used in this study are GaAs/AlGaAs 
heterojunctions grown by molecular beam epitaxy (MBE). Data from six
samples (A through F) will be discussed. Samples A, B and C were
taken from one MBE wafer, D and E from a second, and F
from a third. Each wafer was rotated during growth to ensure high
homogeneity of the electron density $n_s$. These densities (in units of
$10^{11}$ cm$^{-2}$) are close to $n_s = 2.67$ for samples A, B and C;
$n_s = 2.27$ for samples D and E; and $n_s = 1.52$ for sample F. The low
temperature mobility of each is $\mu \ge9 \times 10^6$ cm$^2$/Vs.
Each sample was cleaved (along $\langle$110$\rangle$ directions)
into a $5\times 5$mm square from its parent $\langle$001$\rangle$ wafer.
For samples A,D,E and F, indium contacts were placed
at the corners and the midpoints of the sides of the chip. Hall bar
patterns were lithographically etched onto samples B and C before the
contacts were made. The samples were briefly illuminated at low
temperature with a red LED. Electrical transport measurements were
performed using 2 - 20nA, 13Hz excitation, although for the non-linearity
studies an additional dc current was imposed.

Fig.~1 shows the resistivity\cite{rho} $\rho_{xx}$ at
T=150mK of sample A. Shubnikov-deHaas (SdH)
oscillations commence at around $B$ $\approx$ 60mT and the spin splitting of
the Landau levels is evident by 130mT. The smallness of these fields
attests to the high quality of the 2DEG in this sample.
Between about $B$=5.5 and 11T the Fermi level is in the upper spin branch
of the N=0 lowest LL. This corresponds to Landau level filling fractions
$\nu=h n_s/e B$ between $\nu$=2 and $\nu$=1. Clear signatures of
several fractional quantized Hall states are evident. 
Between $B$=2.8 and 5.5T the Fermi level is in the N=1 second LL
and the developing $\nu$=5/2 FQHE is indicated in the figure. At lower
temperatures this state, and the nearby $\nu$=7/3 state,
strengthen. In the upper spin branch of this same LL a minimum in
$\rho_{xx}$ is seen at $\nu$=7/2 but it does not deepen as the
temperature is reduced.

Below $B$ $\approx$ 2.8T the Fermi level is in the N=2 and higher LLs. 
The inset to Fig.~1 reveals that in both spin branches of the N=2 LL
there are a number of maxima and minima in $\rho_{xx}$. Although
this complex structure does not appear
to be associated with a FQHE (no quantized plateaus in the Hall
resistivity $\rho_{xy}$ have been found), its existence strongly
indicates that electron correlations are important. In the standard
model\cite{standard}, if disorder overwhelms interactions,
a simple peak in the resistivity separates the broad zeroes of adjacent
integer quantum Hall states. 

\begin{figure}[t]
\begin{center}
\epsfxsize=3.3in
\epsffile[56 158 534 500]{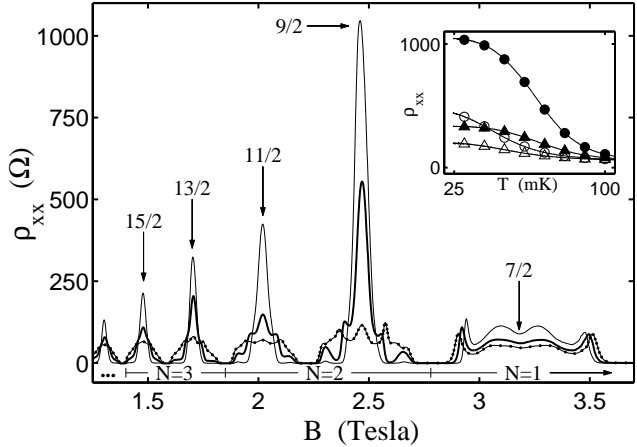}
\end{center}
\caption[figure 2]{Peaks in $\rho_{xx}$ in sample A developing at low
temperatures in high LLs (dotted line: T=100mK; thick line: 65mK; thin
line: 25mK). Inset: temperature dependence of peak height at $\nu$=9/2
(closed circles), 11/2 (open circles), 13/2 (closed triangles) and 15/2
(open triangles).}
\end{figure}

Fig.~2 displays the temperature dependence of the $\rho_{xx}$ features
in the N=1, N=2, and N=3 LLs in sample A. 
Below T=150mK peaks develop at $\nu$=9/2, 11/2, 13/2 and 15/2 which grow
rapidly below 100mK; by 25mK the peak at $\nu$=9/2 has exceeded
1000$\Omega$. Surprisingly, the peaks do not narrow as T is reduced. The
subsidiary structures flanking the peaks do fall with temperature but
are not simply ``consumed'' by the widening of the nearby integer QHE
states; even at 25mK they are still evident\cite{plateau}. This behavior
does not fit the standard model of disorder-driven integer QHE
transitions, but suggests instead that interactions remain
important down to very low T. The observed behavior of
$\rho_{xx}$ in the N=1 LL (only the upper spin branch,
4$>$$\nu$$>$3, is shown in the figure) is qualitatively different. 
Instead of peaks,
there are minima at $\nu$=7/2 and at the fragile
fractional QHE state at $\nu$=5/2. The closeness in magnetic field
of the $\nu$=9/2 and 7/2 filling factors makes this difference
particularly striking. This basic result is just as clearly evident in
samples B through F. In the inset to Fig.~2, $\rho_{xx}$ at $\nu$=9/2,
11/2, 13/2 and 15/2 is plotted versus temperature. 

\begin{figure}
\begin{center}
\epsfxsize=3.3in
\epsffile[56 158 534 500]{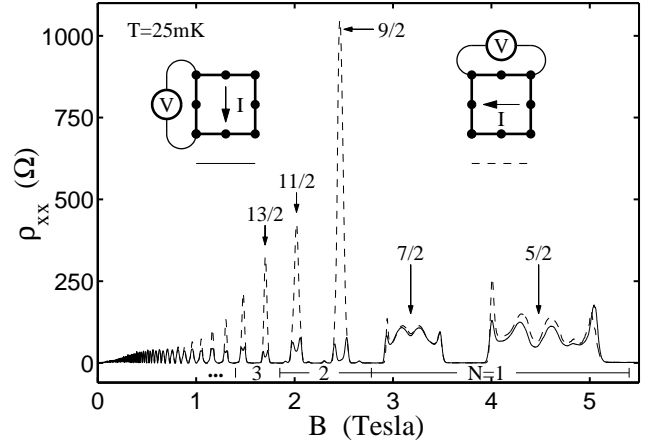}
\end{center}
\caption[figure 3]{Anisotropy of $\rho_{xx}$ in sample A at T=25mK. The
two traces result from simply changing the direction of current
through the sample; the sample itself is {\it not} rotated.}
\end{figure}

Fig.~3 displays our most remarkable finding. The two traces are
resistances measured in sample A at T=25mK for two perpendicular
directions of the current flow through the sample. The diagrams in the
figure depict the difference between the two configurations. 
This seemingly innocuous change vastly alters the
resistivity in the N=2 and several higher LLs. Note that the
solid curve has been multiplied by a factor of 0.62 in order to
match the data sets in the very low field regime. A variation of this
size in the resistance of a quantum Hall sample is quite common and may
simply reflect irregularities in the positions of the contacts. In any
case, this factor in no way obscures the dramatic anisotropy of the
resistivity near the centers of both spin branches of the N=2 through
N=5 LLs. While peaks in $\rho_{xx}$ are evident in one configuration,
relatively deep minima are seen in the other\cite{anisotropy}. At
$\nu$=9/2 the ratio of the resistances is close to 100. Equally striking
is the fact that no comparable anisotropy is apparent in the N=1 LL (nor
the N=0 level which is not shown in the figure). The
anisotropy we are reporting is apparently confined to the centers of the
LLs; well away from half-filling (of each spin branch) the two
resistances again roughly match (after the scaling factor of 0.62 is
applied).  
As the temperature is increased, the anisotropy in $\rho_{xx}$ subsides
until by T=150mK it is no longer significant. 
We emphasize that for no current configuration are any plateaus seen in
the Hall resistance $\rho_{xy}$ at half filling of the N $\ge$ 2 Landau
levels.

\begin{figure}
\begin{center}
\epsfxsize=3.3in
\epsffile[40 162 536 392]{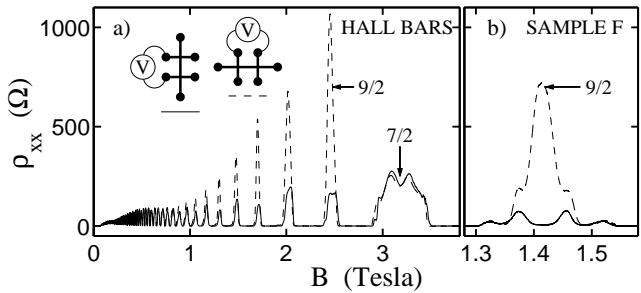}
\end{center}
\caption[figure 4]{a.) Anisotropy of $\rho_{xx}$ at T=25mK measured
using Hall bar samples B (dashed) and C (solid). b.)
Anisotropy of $\rho_{xx}$ at T=15mK measured in the low density sample
F. Current flow configurations as in Fig. 3.}
\end{figure}

By measuring $\rho_{xx}$
with various contact configurations we have determined that the
orientation of the anisotropy is fixed within the sample and is
insensitive to reversal of the magnetic field and thermal cycling to 
room temperature.
Although the geometry of sample A is quite open and the precise current
distribution is unknown, our data suggest that the ``principal axes'' of
the anisotropy are roughly parallel to the sides of the chip.
In order to better define the current path,
Hall bar samples B and C were examined. In
sample B the bar axis was oriented parallel to the
$\langle 1\overline{1}0 \rangle$ crystal axis
while in sample C the bar was aligned along $\langle 110 \rangle$.
Figure 4a compares $\rho_{xx}$ data from these two samples. At 
$\nu$=9/2 the two bars exhibit a large (roughly six-fold)
anisotropy in $\rho_{xx}$ while at $\nu$=7/2 in the N=1 LL the observed
$\rho_{xx}$ is isotropic.
(The data from Hall bar C was multiplied by a factor of 0.75 to match
the amplitude of its SdH oscillations at very low field with those of
Hall bar B.) 
Though not shown in the figure, isotropy is
also observed at $\nu$=5/2 and 3/2. As in sample A, substantial
anisotropy exists near $\nu$=11/2, 13/2, 15/2, and 17/2 before isotropy
returns at low magnetic field. The orientation of the anisotropy in
sample A agrees with that seen in the Hall bars. While
the magnitude of the $\nu$=9/2 anisotropy in the Hall bars is weaker
than that seen in the sample A, it seems clear that the same basic
effect is at work. Finally, the anisotropy effect is also seen in the
lower density samples D and F which come from different MBE wafers.
Figure 4b shows data from sample F in which the 9/2 anisotropy is nearly
100-fold even after {\it magnifying} the data from the 
$\rho_{xx}$-minimum 
configuration by a factor of 1.6 to match the low field SdH oscillations.

The sudden development at very low temperatures of strong temperature
dependences and large anisotropies of the resistivity of clean 2DEGs in
the third and several higher Landau levels suggests that some previously
unappreciated physics is at work. The anisotropy is 
intriguing as it is not at all clear what breaks the in-plane symmetry
of the system. There are various extrinsic effects related to the MBE
growth that might pick out a direction in the 2D plane. A wafer-scale
gradient in the electron density due to the off-normal positions of the
various elemental sources in the MBE chamber is one possibility.
However, our samples were {\it rotated} during their growth specifically
to minimize such gradients. 
Indeed, measurements of the SdH periodicity using many different
voltage and current configurations (including those which produce the large
anisotropy displayed in Fig. 3), show no more than an 0.3\% variation.  This
result strongly suggests that there is at most a very small density gradient.
At the same time, the low onset field of the SdH oscillations
($B$ $\approx$ 60mT, corresponding to Landau filling fraction 
$\nu$ $\approx$ 180) seen in all contact configurations suggests that
density fluctuations on short distance scales are also small.
Among other possible extrinsic symmetry-breaking effects, we mention the
often observed ``slip lines'' which are believed to be steps on the
surface of MBE wafers, and the possibility that the $\langle$001$\rangle$
GaAs substrate was slightly miscut from its parent boule. 

It is possible that the observed anisotropy in $\rho_{xx}$ is due to
some unforeseen order in the static disorder potential. 
This is unlikely since the anisotropy is only seen in the N=2 and a
handful of higher LLs. It is not present in the lowest or first excited
LL nor is it seen in the semiclassical regime well below $B$=1T. The effects
reported here are, however, reminiscent of recently observed transport 
anisotropies near $\nu$=1/2 in the lowest LL in samples where a periodic 
density modulation has been {\it externally} imposed\cite{willett,smet}.
A theoretical basis for understanding this
effect has been proposed\cite{stern}, and it may turn out to be useful in
the present context\cite{halperin} although it would not identify the
{\it source} of the inhomogeneity.

Recently, Koulakov, Fogler and Shklovskii\cite{CDW1} and
subsequently Moessner and Chalker\cite{CDW2} have proposed that in
a clean 2DEG in the N=2 and higher LLs the uniform electron liquid may
be unstable against the formation of charge density waves (CDW). They
further suggest that near half-filling of the
LL the CDW is a unidirectional ``stripe phase'' having a wavelength of 
order the cyclotron radius.
In this stripe phase the electron density in the uppermost LL alternates
between zero and full filling. At $\nu$=9/2 this implies there are
stripes of the incompressible QHE states $\nu$=4 and $\nu$=5.
While it is surely
plausible that electrical transport in such a unidirectional phase would
be anisotropic, it is not clear what would pin the stripes or why they are
apparently coherent over the macroscopic size of our samples.

\begin{figure}
\begin{center}
\epsfxsize=3.3in
\epsffile[60 148 520 502]{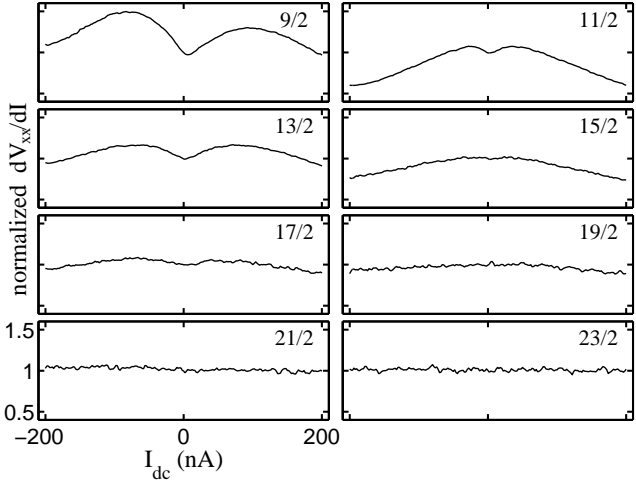}
\end{center}
\caption[figure 5]{Non-linearity of differential resistivity
dV$_{xx}$/dI in sample C at half-filling of several high LLs at T=25mK.
The resistivity is normalized by its value at I$_{dc}$=0. In each panel
I$_{dc}$ runs from $-200$ to $+200$nA and the normalized dV$_{xx}$/dI
from 0.5 to 1.5.}
\end{figure}

To further investigate magneto-transport in highly excited LLs, we have
examined the linearity of $\rho_{xx}$. Fig.~5 summarizes our T=25mK
results from sample E. The figure displays the differential resistivity
dV$_{xx}$/dI at $\nu$=9/2, 11/2, etc. measured using 5nA, 13Hz
excitation, in the presence of an added dc current I$_{dc}$. The data
were taken using a contact configuration for which the resistivity (i.e.
dV$_{xx}$/dI at I$_{dc}$=0) exhibits strong peaks at these filling
factors. In each panel I$_{dc}$ runs from $-200$ to +200nA, and the
plotted dV$_{xx}$/dI data are normalized by their value at I$_{dc}$=0.
Several of the data sets show marked non-linearity. At $\nu$=9/2,
where the effect is strongest, dV$_{xx}$/dI at first increases
substantially as I$_{dc}$ is applied. At about $\pm$100nA it
reaches a maximum and then falls off at higher current. Since
$\rho_{xx}$ at $\nu$=9/2 falls with rising temperature, it is clear that
the initial rise in dV$_{xx}$/dI at small I$_{dc}$ is {\it not
consistent with electron heating}. At large I$_{dc}$ the falling of
dV$_{xx}$/dI may indeed be due to heating. 
Qualitatively similar non-linearities are
seen at half filling of the higher LLs, but the effect eventually dies
away at low magnetic field. Note the oscillatory strength of the
non-linearity: strong at 9/2, weaker at 11/2, relatively stronger at
13/2, weak again at 15/2, and so on. As with the anisotropy, the
non-linearity is a low temperature phenomenon; by 150 mK it is
essentially gone. It is more fragile than the anisotropy; while clearly
evident in some samples it is weaker in others. We emphasize that in
each LL the non-linearity is seen over a range of $\nu$ around 
half-filling. Finally, although weak non-linearities have sometimes been
seen in the N=1 and N=0 LLs, they appear qualitatively different from
that seen in the N=2 and higher LLs.

The increasing differential resistivity for small dc currents also
implies an increasing {\it conductivity}. This follows from inverting
the resistivity tensor and noting that the Hall resistivity $\rho_{xy}$
is several times larger than $\rho_{xx}$ over the field range of
interest. Although an increasing conductivity is suggestive of a
depinning mechanism, we have seen no evidence of a sharp threshold; the
non-linearity appears to turn on continuously.

In conclusion, we have reported dramatic magneto-transport anomalies
that are specific to the third and higher Landau levels in clean
2DEG's in GaAs/AlGaAs heterostructures. These effects, which only appear
at low temperature, include huge anisotropies in the
resistivity near half filling of the spin-resolved Landau levels as well
as intriguing non-linearities. The origin of these effects is unclear 
but the data are consistent with the spontaneous development of
an anisotropic electronic configuration. The occurrence of the same
basic transport signatures in several adjacent highly excited Landau
levels points to a generic mechanism.

We thank J.~Chalker, S.~M. Girvin, B.~Halperin, A.~H. MacDonald, and 
K.~Yang for helpful discussions and Ian Spielman for help in 
determining the crystallographic axes in our samples. 
This work was supported by the National 
Science Foundation.


\begin{references}
\bibitem{rho} As is commonplace, we employ the term resistivity even
though the actual measured quantity is resistance; in square samples
the factor relating the two is of order unity.
\end{references}
\end{document}